\documentclass[11pt, a4paper]{article}

\usepackage[utf8]{inputenc}
\usepackage[T1]{fontenc}
\usepackage[margin=1in]{geometry} 

\usepackage{amsmath, amssymb, amsfonts}
\usepackage{graphicx}
\usepackage{dcolumn}
\usepackage{bm}
\usepackage{xcolor}
\usepackage{soul}
\usepackage{balance}

\usepackage{cite} 

\usepackage[colorlinks=true, linkcolor=blue, citecolor=blue, urlcolor=blue]{hyperref}

\usepackage{authblk} 

\def\be{\begin{equation}}
\def\ee{\end{equation}}
\def\ba{\begin{eqnarray}}
\def\ea{\end{eqnarray}}

\def\bl#1\el{\begin{align}#1\end{align}}

\usepackage{authblk} 

\title{\textbf{Time Crystals in Coupled Exciton-Polariton Condensates}}
\author[1]{Xuan Ye}
\author[2]{Hong-Jin Xiong}
\author[3,4,5]{Alexey Kavokin}
\author[1,6]{Sanjib Ghosh}

\affil[1]{School of Science and Engineering, The Chinese University of Hong Kong (Shenzhen), Longgang, Shenzhen, Guangdong 518172, P.R. China}

\affil[2]{Department of Applied Physics, The Hong Kong Polytechnic University, Hong Kong, China}

\affil[3]{Abrikosov Center for Theoretical Physics, Moscow Center for Advanced Studies, Moscow 141701, Russia}

\affil[4]{School of Science, Westlake University, 18 Shilongshan Road, Hangzhou 310024, Zhejiang Province, China}

\affil[5]{Department of Physics, St. Petersburg State University, University Embankment, 7/9, St. Petersburg, 199034, Russia}

\affil[6]{Beijing Academy of Quantum Information Sciences, Beijing, 100193, P. R. China}

\date{\today}

\begin{document}

\maketitle

\begin{abstract}
In this paper, we show that a time crystal can emerge in coupled exciton-polariton condensates without periodic external driving, enabled instead by incoherent gain and dissipation channels inherent to semiconductor microcavities. We present a full quantum description of these processes that recovers the established effective theory at the mean-field level. We analytically determine the mean-field phase diagram for the time-crystalline phase and find that its emergence requires the ratio of Kerr nonlinearity to nonlinear dissipation to exceed $\sqrt{5/4}$. Within this regime, the periodic oscillation of the particle numbers forms an attractor that is insensitive to the initial conditions. Numerical bifurcation diagrams reveal transitions between the time-crystalline phase and various steady phases, in excellent agreement with the analytical results.
Using Bogoliubov perturbation theory, we evaluate the leading-order quantum corrections and find that, over a wide parameter range, these corrections remain periodic and much smaller than the mean-field background, thereby establishing the robustness of the time crystal.
\end{abstract}

\section{Introduction}
A time crystal is the temporal analog of a spatial crystal, characterized by spontaneous breaking of time translation symmetry. Wilczek originally proposed that a time crystal could be realized by turning a steady system into one that moves periodically in its ground state \cite{Frank1,Frank2,Wilczek3,Li2012}. However, no-go theorems later showed that such a phase cannot exist in equilibrium \cite{Bruno,Watanable}. This result motivated studies of periodically driven systems, known as discrete or Floquet time crystals \cite{Sacha, Gong, Yao, Yang, Else, Huang,Pizzi,Liu,Yao2}. These time crystals have been observed experimentally in superfluids \cite{Smits1,Autti1}, superconducting qubits and quantum processors \cite{Frey2022}, trapped ions \cite{Zhang2017,Kyprianidis2021}, and nuclear-spin systems \cite{Choi2017,Rovny2018,Pal2018,OSullivan2020}.

In recent years, the realization of time crystals in dissipative, periodically driven  quantum systems has attracted significant attention \cite{Sacha2018,Lled,Cabot}.
Such phenomena manifest themselves in many-body spin systems \cite{Tucker2018,Iemini2018,Gambetta2019,Buca2019} and 
in open nonlinear photonic cavities with external driving~\cite{Seibold,Bakker,Li2024}. In the latter, 
\begin{figure}[htp!]
\centering
\includegraphics[width=0.75\linewidth]{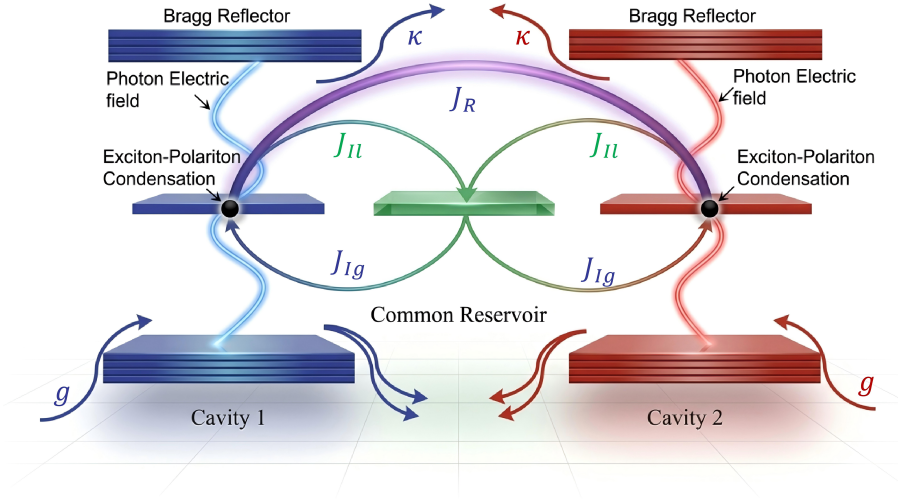}
\caption{\textbf{Schematic of the coupled microcavity setup.} Two microcavity exciton-polariton condensates are coupled to a common exciton reservoir. Each condensate undergoes local linear gain and dissipation at rates $g$ and $\kappa$, respectively, alongside nonlinear loss at rate $\eta_I$. The shared reservoir mediates correlated-incoherent gain and dissipation with rates $J_{Ig}$ and $J_{Il}$, while coherent quantum tunneling between the cavities occurs at rate $J_R$.}
\label{Fig00}
\end{figure}
the periodic modulation is often provided by a constant coherent field, which acts as a periodic driving when transformed from the rotating to the laboratory frame.
At the mean-field level, these systems act as driven self-sustained Van der Pol (VdP) oscillators \cite{Lee1,Niels,Walter}, where the particle number oscillates and shows time-crystalline behavior. 
In this context, coupled cavities
, both with and without nonlinear dissipation, have been explored \cite{Seibold,Bakker}. Alternatively, incoherent reservoir dynamics can be used to realize the time-crystalline phase~\cite{Nalitov2019}. Complementary studies have examined quantum fluctuations in VdP oscillators \cite{Weiss, Navarrete, Sonar2018, Benlloch}, alongside the driven dynamics of a single nonlinear cavity subject to gain and dissipation within a full quantum framework~\cite{Li2024}. Although these studies suggest that time crystals can exist in photonic cavities, their characterization as classical self-sustained oscillations remains a point of contention~\cite{NavarreteBenlloch2024}.
It is natural to ask whether  time-crystalline order can emerge in the absence of classical time-dependent mechanisms, such as periodic external driving or a dynamic reservoir.




In this paper, we show that a time crystal can emerge in coupled exciton-polariton condensates without periodic external driving or a dynamic reservoir, enabled instead by time-independent gain and dissipation channels inherent to semiconductor microcavities. Beyond the mean-field level, we formulate the corresponding quantum theory using the Lindblad master equation. Under the mean-field approximation, our quantum description recovers the effective theory previously developed for EP condensates \cite{Berloff2017} and verified experimentally \cite{Berloff2017Exp}. In this regime, the system dynamics are essentially governed by the coupled Stuart–Landau (SL) equations \cite{Selivanov,Ku,Mendola,Millan}.
To map out the phase diagram for the time-crystalline state, we identify the parameter regime where all fixed points are unstable. These fixed points are analytically derived from the SL equations and their stability is assessed via Jacobian analysis \cite{Khalil2002}.
In this regime, the particle numbers oscillate periodically at long times and are insensitive to the initial state. 
Based on our full quantum description, we evaluate the quantum fluctuations using Bogoliubov perturbation theory \cite{PitaevskiiStringari2003}. The closed covariance matrix equation for the second-order moments of the bosonic fluctuation operators is derived. Solving this equation numerically shows that, within the time-crystalline parameter regime determined by the SL equations, the leading-order quantum corrections oscillate stably and remain much smaller than the background particle numbers over a wide range of parameters, indicating that the time crystal persists even when quantum effects are included.

\section{Physical Model and Fixed Points}
Consider the configuration of coupled EP condensates in semiconductor microcavities shown in Fig. \ref{Fig00}. We introduce the master equation    
$(\hbar=1)$
\bl
\frac{d}{dt}\hat{\rho}=&-i[\hat{H},\hat{\rho}]
+\sum_{i=1}^2\big( \kappa \hat{\cal D}[\hat{a}_i]\hat{\rho} +g \hat{\cal D}[\hat{a}_i^\dag]\hat{\rho}+ \eta_I  \hat{\cal D}[\hat{a}_i^2]\hat{\rho}\big)+J_{Il}\hat{\cal D}[\hat{a}_1+\hat{a}_2]\hat{\rho}+J_{Ig}\hat{\cal D}[\hat{a}_1^\dag+\hat{a}_2^{\dag}]\hat{\rho},\label{Lindbladequatio}
\el
where $\hat{\rho}$ is the density matrix, $\hat{a}_i$ ($\hat{a}_i^\dag$) is the bosonic annihilation (creation) operator for $i$th- species ($i = 1, 2$). The Liouvillian superoperator for an operator $\hat{o}$ is defined 
as $\hat{\cal D}[\hat{o}]\hat{\rho}\equiv \hat{o}\hat{\rho}\hat{o}^\dag-\frac12\{\hat{o}^\dag\hat{o},\hat{\rho}\}$. 
We include the local linear gain (dissipation) $g$ ($\kappa$) and nonlinear dissipation $\eta_I$, which are determined by statistical independent reservoirs \cite{Lled,Leghtas2015}. We introduce correlated gain ($J_{Ig}$) and dissipation ($J_{Ik}$) terms to describe the incoherent coupling of the condensates to a common reservoir, which accuratly reproduces the established effective theory \cite{Berloff2017, Rubo2012}.
The Hamiltonian is given by,
\begin{equation}
\hat{H}=\sum_{i=1}^{2}(\omega \hat{a}_i^{\dag}
\hat{a}_i+\frac{\eta_R}{2}\hat{a}_i^{\dag}\hat{a}_i^{\dag}\hat{a}_i\hat{a}_i)+J_R (\hat{a}_1^{\dag}\hat{a}_2+\hat{a}_2^{\dag}\hat{a}_1),
\end{equation}
where $\omega$ is the energy of an EP condensate. $\eta_R$ represents the Kerr nonlinearity coefficient, and $J_R$ is the quantum tunneling magnitude \cite{Rubo2012}. The Hamiltonian is subject to {\it neither periodic nor coherent driving}.
Here, all parameters are real and positive, and are expressed in units of $\kappa_0 = 10^2 \,\mathrm{GHz}$. 
 
An EP condensate, typically comprising $10^2\sim10^4$ particles \cite{Jiepeng}, are well-described within the mean-field approximation. In this regime, the annihilation operator can be written as $\hat{a}_i = \alpha_i+\delta \hat{a}_i$, where $\alpha_i=\langle \hat{a}_i \rangle$, and $\langle...\rangle$ denotes the expectation value. The quantum fluctuation part vanishes on average $\langle\delta\hat{a}_i\rangle=0$.
Equation \eqref{Lindbladequatio} then reduces to the coupled SL equations \cite{Selivanov,Ku,Mendola,Millan}
\begin{equation}
\dot{\alpha}_i = \big(p - i\omega
 -\tilde{\eta}|\alpha_i|^2 \big)\alpha_i
 +(J_I - i J_R) \sum_{j\neq i}\alpha_j,
 \label{modeu1}  
\end{equation}
where the overdot denotes the time derivative, 
\begin{figure}[htbp]
  \centering  \includegraphics[width=0.75\linewidth]{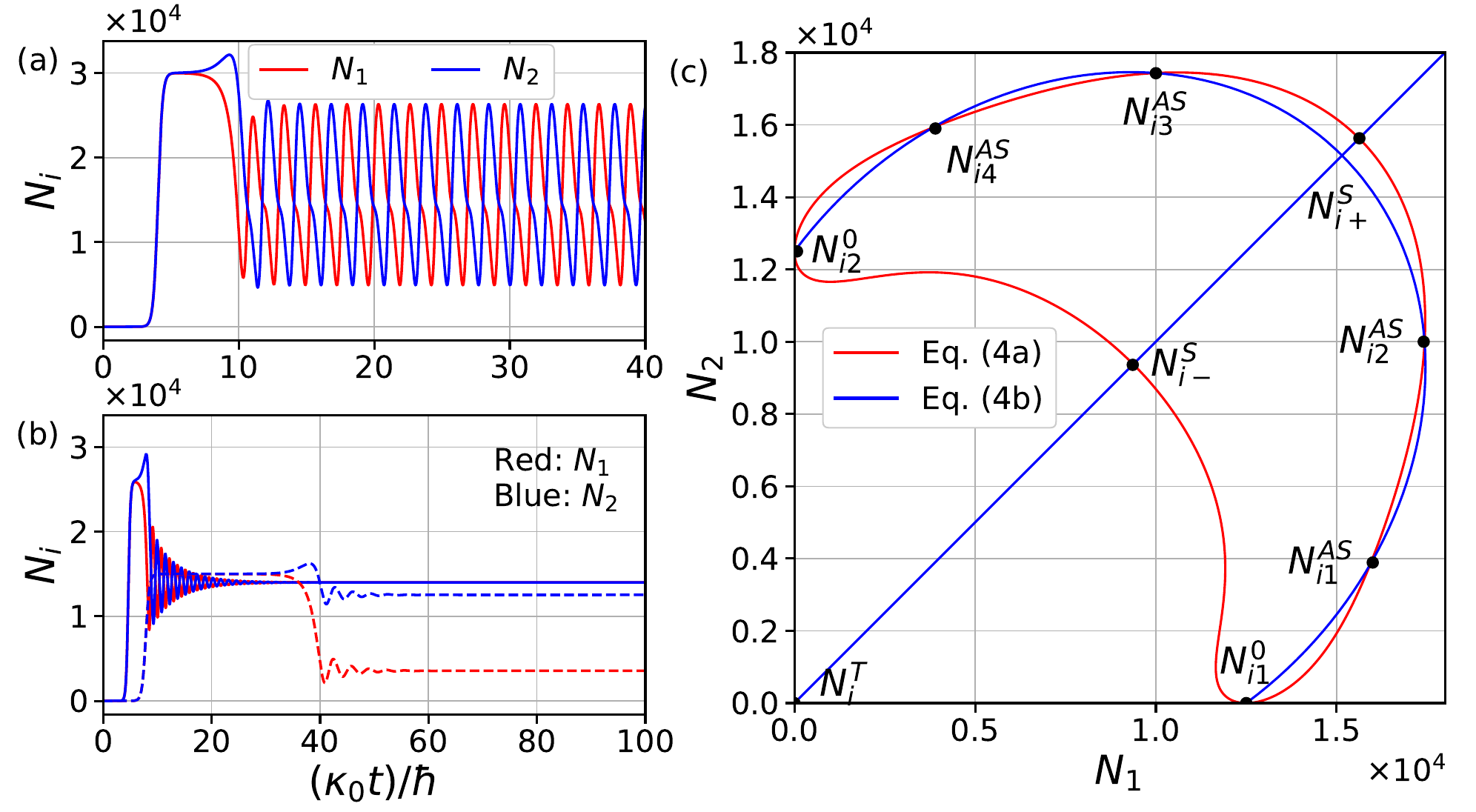}
\caption{\textbf{Time-crystalline and steady states of the system.} (a) In the time-crystalline phase, the particle numbers oscillate persistently in the long-time regime for a representative parameter set ${\bf P}/\kappa_0=(2, 3\times10^{-4}, 10^{-4}, 1, 1)$.
(b) Steady-state behavior outside the time-crystalline region. For parameter set
${\bf P}/\kappa_0=(2, 3\times10^{-4}, 10^{-4}, 1, 0.6)$, both particle numbers converge to a common value, representing a symmetric steady state. In contrast, for ${\bf P}/\kappa_0=(0.5, 3\times10^{-4}, 10^{-4}, 1, 1)$, the populations converge to different values.
(c) Fixed points determined by Eqs. \eqref{Nequation} for
${\bf P}/\kappa_0=(4, 3\times10^{-4}, 3.2\times10^{-4}, 2, 1)$.
The initial conditions $R_i=10^{-3}$ and $\phi_i=0.2\pi$ are used in solving mean-field equations throughout this paper.
}
 \label{Figm1}
\end{figure}
$\tilde{\eta} =\eta_I + i\eta_R$ is the nonlinear coefficient, $J_I \equiv (J_{Ig}-J_{Il})/2$ is the  net gain rate due to the correlated processes, and 
$p\equiv (g-\kappa+J_{Ig}-J_{Il})/2$ is the net gain rate arising from both local and correlated processes. 
Here, we consider a gain dominated system with $p>0$ and $J_I>0$. 


Substituting the variables $\alpha_i=R_ie^{i\theta_i}$ into Eq. \eqref{modeu1} yields a set of three coupled equations, 
\begin{subequations}\label{eqss}
\bl
\dot R_1 &= R_1(p  - \eta_I R_1^2) + R_2\big( J_I\cos\phi - J_R\sin\phi \big), \label{eq:R1}\\
\dot R_2 &=R_2( p  - \eta_I R_2^2) + R_1\big( J_I\cos\phi + J_R\sin\phi \big), \label{eq:R2}\\
R_1R_2\dot{\phi}&=  -\eta_R (R_1^2-R_2^2) -J_R (R_2^2-R_1^2) \cos\phi- J_I  (R_2^2+R_1^2)  \sin\phi,\label{eq:R3}
\el
\end{subequations}
which are independent of the energy $\omega$, and $\phi\equiv (\theta_1-\theta_2)$ denotes the phase difference. Depending on the parameters and initial conditions, the long-time dynamics of the particle numbers $N_i\equiv |\alpha_i|^2= R_i^2$ typically fall into one of three categories: spontaneous  persistent oscillations (time-crystalline behavior), or convergence to either symmetric (equal values) or   asymmetric(unequal values) steady states, as illustrated in Figs. \ref{Figm1} (a) and (b). 

To identify the time-crystalline regime, one could scan the entire five-dimensional parameter space ${\bf P}\equiv\{p, \eta_R, \eta_I, J_R, J_I\}$; however, such an exhaustive search is computationally expensive. Direct identification is also difficult, as Eqs. \eqref{eqss} admit no analytic solutions in the oscillatory regime. 
Instead, we map the boundaries of the time-crystalline phase  by identifying the regimes where stable steady solutions are absent. We thus conjecture that {\it the time-crystalline region is given by the intersection of the unstable regions of all fixed points.}

In the following, we determine all fixed points of Eqs. \eqref{eqss}. The stability analysis is provided in the following section. Setting $\dot{R}_i=\dot{\phi}=0$ and eliminating  $\phi$ yield two 
algebraic equations 
\begin{subequations}\label{Nequation}
\bl
J_I^2S_p^2D^2+J_R^2 (p S-\eta_I S_M)^2&=4 J_I^2 J_R^2 M,\label{Nequation1}
\\
[J_R^2 (\eta_I S_M-p S)+J_I ^2S_pS]D&=-2\eta_R J_I J_R M D,\label{Nequation2}
\el
\end{subequations}
where $S\equiv N_1+N_2$, $D\equiv N_1-N_2$, 
$M\equiv N_1 N_2$, $S_p\equiv (\eta_I S-p )$ and $S_M\equiv (S^2-2M)$. 
For $D=0$, Eq. \eqref{Nequation2} becomes an identity, and Eq. \eqref{Nequation1} consequently yields a pair of symmetric fixed points,
\bl
N_{1\pm}^S=N_{2\pm}^{\text{S}}=\frac{p\pm J_I}{\eta_I},\label{Rpm}
\el
corresponding to the in-phase ($\phi_+=0$) and anti-phase ($\phi_-=\pi$) states \cite{ARONSON1990403}. In the Supplemental Material \cite{supp1}, we show that Eqs. \eqref{Nequation} admit four additional   asymmetricsolutions $(N_{i1}^{\text {AS}}, N_{i2}^{\text{AS}}, N_{i3}^{\text {AS}},N_{i4}^{\text{AS}})$, two single-mode solutions 
$(N_{i1}^{0}, N_{i2}^{0})$, and the trivial solution $N_i^{\text T}=0$. For illustration, Eqs. \eqref{Nequation1} and \eqref{Nequation2} are plotted in Fig. \ref{Figm1} (c), with the solutions indicated by black dots. 
Hereafter, we exclude single-mode solutions, as they are not valid solutions of Eqs. \eqref{eqss}, and  the null-particle state ($N^\text{T}_i$) is of no interest. The system can converge to different fixed points depending on both their stability and the initial conditions. For example, the steady states indicated by solid  and dashed lines in Fig. \ref{Figm1} (b) correspond to the fixed points $N_{i-}^\text{S}$ and 
$N_{i4}^{\text{AS}}$, respectively. 


\section{Parameter Space of Time Crystal}
Next, we employ the Jacobian matrix method \cite{Khalil2002} to identify the intersection of the instability regions for all fixed points, which is defined as the time-crystalline regime. While we focus here on the symmetric fixed points, the analogous stability analysis for   asymmetricstates is provided in the Supplemental Material \cite{supp1}.
Linearizing Eqs. \eqref{eqss} around the symmetric fixed points $N^{\text S}_{i\pm}$ yields
\bl
\delta\dot{\bf R}
=
{\bf J}_{\pm}^{\bf S}\cdot
\delta{\bf R},
\label{variationeq}
\el
where \(\delta{\bf R}\equiv(\delta R_1,\delta R_2,\delta\phi)^{\rm T}\), and \({\bf J}_{\pm}^{\bf S}\) is the Jacobian matrix evaluated at \(N_{i\pm}^{\text{S}}\). For each symmetric fixed point, there are three eigenvalues, denoted by $\left(\lambda_{1\pm}^{\rm S},\lambda_{2\pm}^{\rm S},\lambda_{3\pm}^{\rm S}\right)$, 
whose explicit expressions are given in Ref.~\cite{supp1}. According the linear-stability criterion \cite{Arnold1973}, a symmetric fixed point is unstable if at least one eigenvalue
has a positive real part
\bl
\mathrm{Re}\!\left[\lambda_{i\pm}^{\rm S}\right]>0.
\label{dwedh}
\el
which leads to the exponential growth of the perturbation vector $\delta {\bf R}$. 
\begin{figure}
\centering
\includegraphics[width=0.75 \linewidth]{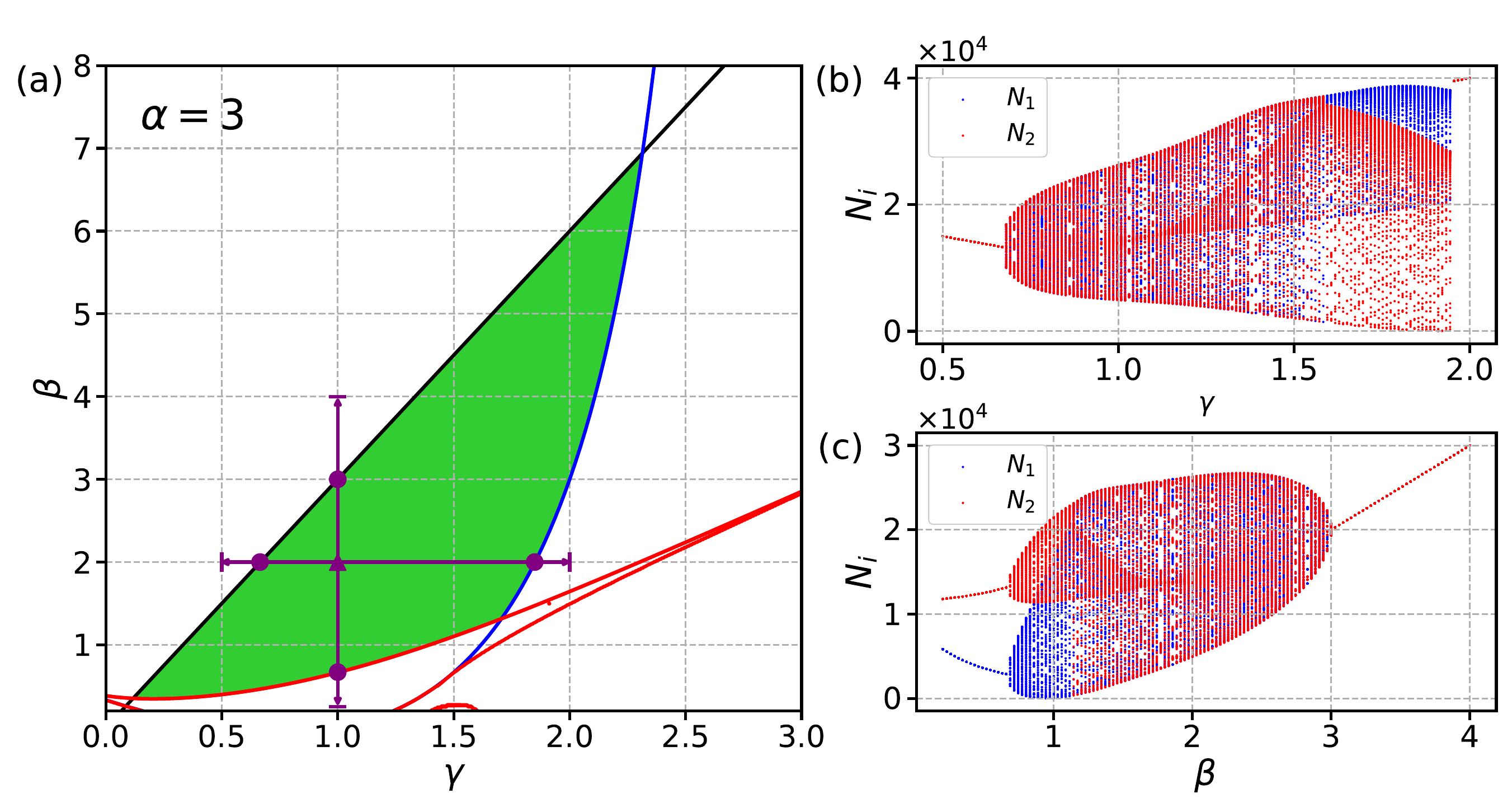}
\caption{\textbf{The phase diagram.} (a) The green shaded region in the \((\gamma,\beta)\) plane identifies the stable time-crystalline phase, with its boundary defined by the intersection of the instability thresholds for the in-phase symmetric (black), anti-phase symmetric (blue), and   asymmetric(red) fixed points.
(b) Bifurcation diagram versus $\gamma$ for $(p, \eta_R, \eta_I, J_R)/\kappa_0=(2, 3\times10^{-4}, 10^{-4}, 1)$, with $J_I/\kappa_0\in[0.5,2.0]$ along the horizontal purple line in (a).
(c) Bifurcation diagram versus $\beta$ for $(\eta_R, \eta_I, J_R, J_I)/\kappa_0=(3\times10^{-4}, 10^{-4}, 1, 1)$, with $p/\kappa_0\in[0.2,4]$ along the vertical purple line in (a).}
\label{Fig0}
\end{figure}

Introducing the variables $\alpha\equiv \eta_R/\eta_I$, $\beta\equiv p/ J_R$, and $\gamma\equiv J_I/ J_R$, the instability regions for the $N_{i+}^{\text S}$ and $N_{i-}^{\text S}$, as determined by \eqref{dwedh}, are given by (See Ref. \cite{supp1} for details)
\begin{subequations}\label{unstableinpahseneww} 
    \bl
\beta&<3\gamma,\label{unstableinpahsenew1s}
\\
\beta\gamma+2\gamma^2 +&1-\alpha (\gamma+\beta)<0.\label{unstableinpahsenew}
\el
\end{subequations}
The boundaries defined by \eqref{unstableinpahsenew1s} and \eqref{unstableinpahsenew} are marked by the black and blue lines in Fig. \ref{Fig0}(a). Parameter sets satisfying both inequalities exclude convergence to equal particle numbers. Convergence to unequal particle numbers is ruled out by choosing parameters in the regime where the   asymmetric fixed points are unstable. The red lines in Fig. \ref{Fig0}(a) show the boundaries determined by the  asymmetric fixed points. For the nonlinearity ratio $\alpha=3$, the green region in Fig. \ref{Fig0}(a) denotes the time-crystalline parameter space. For example, the parameter set ${\bf P}/\kappa_0=(2, 3\times10^{-4}, 10^{-4}, 1, 1)$ yields the point $(\alpha, \beta, \gamma)=(3, 2, 1)$, marked by the purple triangle in Fig. \ref{Fig0}(a). This point exhibits stable oscillations of the particle numbers, see Fig. \ref{Figm1}(a), characteristic of a time crystal. This oscillatory solution forms an attractor and is insensitive to the initial conditions: we have tested initial particle numbers with $N_{1}=N_{2}$ over the range $(10^{-16},10^{4})$ and the initial phase difference $\phi$ over $(\epsilon,\pi-\epsilon)$, with $\epsilon=10^{-9}\pi$. Similar attractor behavior also occurs in polariton superfluids trapped in C-shaped potentials \cite{Sun2024}. The time-crystalline regime is three-dimensional, since it depends only on the relative strengths of the parameters, reflecting a uniform scaling symmetry. 

In a gain-dominated system ($\beta,\gamma>0$), the emergence of a time crystal requires the nonlinearity ratio to exceed the threshold
\bl
\alpha>\sqrt{5/4},
\label{critialalpha}
\el
otherwise the unstable regions defined by  \eqref{unstableinpahseneww} do not overlap, as conformed numerically. The overlap region increases with $\alpha$. Since $\alpha$ is invariant under the transformations $\eta_{(R,I)}\rightarrow \eta_{(R,I)}'=C^{-1}\eta_{(R,I)}$ and $\alpha_{i}\rightarrow \alpha_i' = C \alpha_i $, similar dynamical time-crystal behavior can arise in systems with different particle number, because Eqs. \eqref{modeu1} are invariant under the same transformations.

The transitions among the symmetric, asymmetric, and time-crystalline states are examined numerically through the bifurcation diagrams in Figs. \ref{Fig0} (b) and (c). We scan the parameter space along two perpendicular purple lines shown in Fig. \ref{Fig0} (a). For each parameter set, the particle numbers are sampled after the system reaches its long-time dynamics. 
Fig.~\ref{Fig0} (b) shows the transition between symmetric low- and high- particle-number states, with an intermediate time-crystalline phase emerging for $\gamma \in(0.67,1.85)$. These bifurcation points align precisely with the vertical purple line’s intersections with the instability boundaries in Fig.~\ref{Fig0} (a). Similarly, Fig.~\ref{Fig0} (c) depicts the transition from a   asymmetricto a symmetric state, featuring a time-crystalline interval for \(\beta\in(0.67,3.0)\). The corresponding bifurcations match the horizontal purple line’s intersections in Fig.~\ref{Fig0} (a).

\section{Quantum Fluctuations}
In what follows, we examine quantum effects in the system governed by Eq.~\eqref{Lindbladequatio}. Unlike photonic cavities, whose particle numbers are typically of order $10\sim10^2$ \cite{Li2024}, each EP condensate considered here contains on the order of $10^2\sim10^4$ particles \cite{Jiepeng}. Consequently, solving Eq. \eqref{Lindbladequatio} using full quantum algorithms \cite{qutip5} would, even under optimistic estimates, require truncation of the Fock space at occupations of order $10^4\sim10^6$. For coupled condensates, the Hilbert-space dimension would scale as $10^8\sim10^{12}$, making numerical integration intractable. We therefore focus on the leading-order quantum corrections to the particle numbers, obtained from Bogoliubov perturbation theory \cite{PitaevskiiStringari2003}.

\begin{figure}[!t]
\centering
\includegraphics[width=0.75\linewidth]{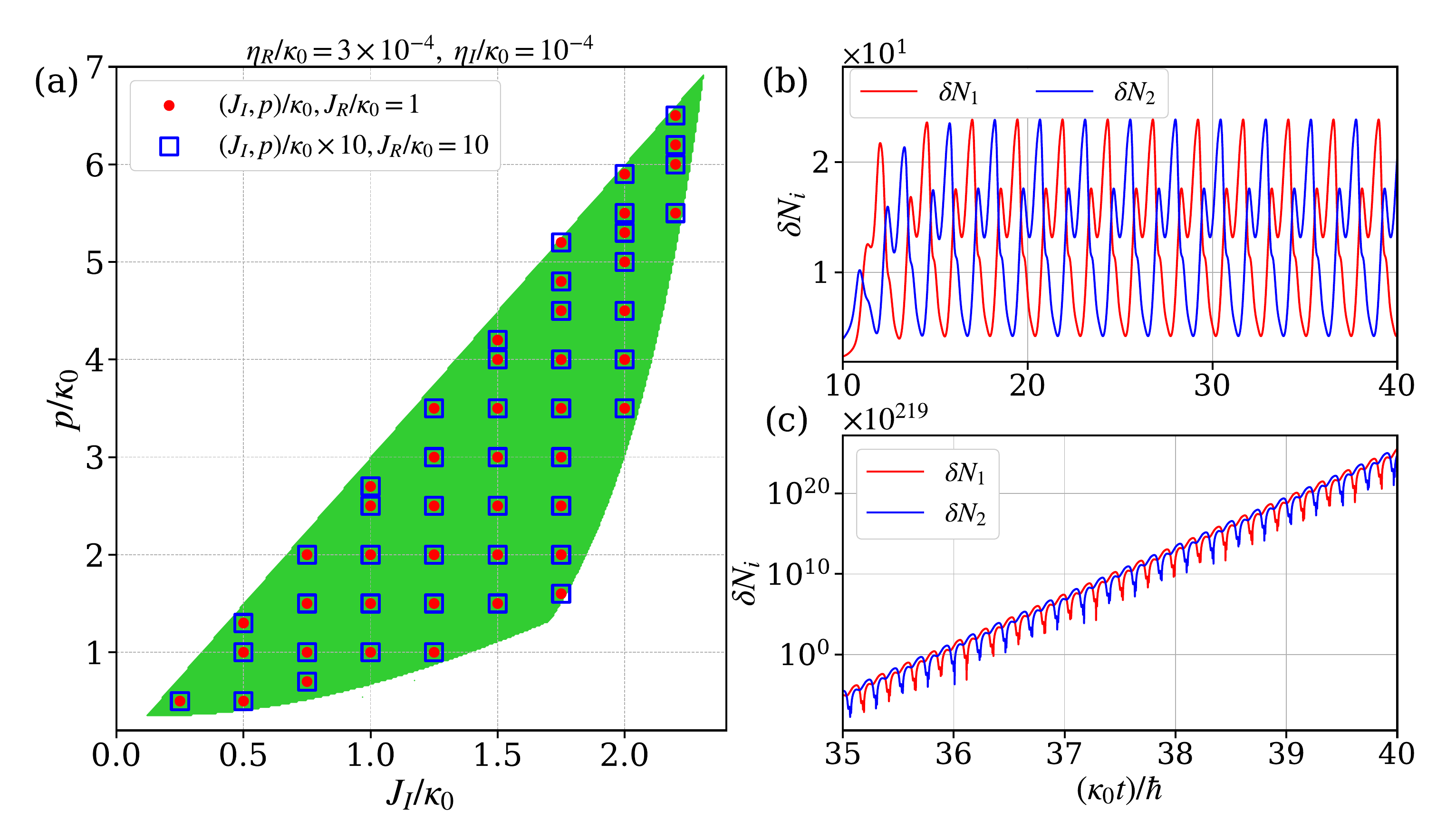}
\caption{\textbf{Robustness against quantum fluctuations.} (a) Green region: time-crystalline phase determined from the mean-field equations. Red dots: parameter sets yielding stable quantum oscillations. Blue squares: parameter sets for which the quantum fluctuations grow exponentially, indicating the breakdown of perturbation theory.
(b) Stable quantum oscillations for
$(\eta_R,\eta_I,J_R,\omega,g_{Ig},g_{Il},g,\kappa)/\kappa_0
=(3\times10^{-4},10^{-4},1,10^{4},3,1,3,1)$.
(c) Exponentially growing quantum fluctuations for
$(\eta_R,\eta_I,J_R,\omega,g_{Ig},g_{Il},g,\kappa)/\kappa_0
=(3\times10^{-4},10^{-4},10,10^{4},30,10,30,10)$.}
\label{Fig4}
\end{figure}

Transforming Eq.~\eqref{Lindbladequatio} into the Heisenberg equation form and expanding $\hat{a}_i\approx \alpha_{i}+\delta\hat{a}_i$, we   retain terms up to second order in $\delta \hat{a}_i$ and $\delta \hat{a}_i^\dag$. This yields a closed set of sixteen equations for the second order moments
$\langle\delta\hat{a}_i\delta\hat{a}_j\rangle$, $\langle\delta\hat{a}_i^\dag\delta\hat{a}_j^\dag\rangle$, $\langle\delta\hat{a}_i^\dag\delta\hat{a}_j\rangle$, and $\langle\delta\hat{a}_i\delta\hat{a}_j^\dag\rangle$. The covariance matrix then obeys (see Ref. \cite{supp1} for details.)
\begin{equation}
\dot{{\bf V}} = {\bf A} {\bf V} + {\bf V} {\bf A}^\dag + {\bf D}, \label{coveq}
\end{equation}
where ${\bf V}\equiv \langle {\bf v}^\dag{\bf v}\rangle$ and  ${\bf v}\equiv (\delta \hat{a}_1, \delta \hat{a}_2, \delta \hat{a}_1^\dag, \delta \hat{a}_2^\dag)^{\text{T}}$. The drift and diffusion matrices are 
\begin{equation}
{\bf A}\equiv\begin{pmatrix}
X & Y \\
Y^\dag & X^\dag
\end{pmatrix}, \quad
{\bf D}\equiv\begin{pmatrix}
G & 0 \\
0 & L
\end{pmatrix}. \label{Liponuov}
\end{equation}
with $2 \times 2$ blocks defined as $X=[p+i\omega-(\eta_I-i\eta_R)S]I+(J_I+i J_R)\sigma_x-(\eta_I-i\eta_R)D\sigma_z$, $Y=-\frac{1}{2}(\eta_I-i\eta_R)[(\alpha_1^{*2}+\alpha_2^{*2})I+(\alpha_1^{*2}-\alpha_2^{*2})\sigma_z]$, $G=(g+J_{lg})I + J_{lg}\sigma_x$, and $L=(\kappa+J_{ll}+2\eta_I S)I + J_{Il}\sigma_x+2\eta_I D\sigma_z$, where $I$ is the identity matrix and $\sigma_{x,z}$ are  Pauli matrices. The diffusion matrix ${\bf D}$ describes vacuum quantum noise, which  seeds fluctuation 
dynamics  even in the absence of initial fluctuations; it  consists of a constant matrix $G$ associated with gain and a time-dependent matrix $L$ associated with dissipation.

The perturbed particle numbers in Eq.~\eqref{coveq} depend on eight independent parameters, in contrast to the background ones, which are determined by five parameters through Eqs.~\eqref{eqss}. Consequently, the mapping from the full parameter space to the time-crystalline space \((\alpha,\beta,\gamma)\) is many to one. To demonstrate the existence of parameter sets for which the quantum fluctuations remain periodic and small, we scan the green region in Fig.~\ref{Fig0}(a) with the initial condition \({\bf V}=0\). Fixing \((\eta_R,\eta_I,J_R)/\kappa_0=(3\times10^{-4},10^{-4},1)\), we find that parameter sets \((J_I,p)/\kappa_0\) of order \(1\), marked by red dots, yield long-time stable oscillatory fluctuations in the particle numbers \(\delta N_i\equiv\langle\delta\hat{a}_i^\dagger\delta\hat{a}_i\rangle\), as shown in Fig.~\ref{Fig4}(b). By contrast, for \((\eta_R,\eta_I,J_R)/\kappa_0=(3\times10^{-4},10^{-4},10)\), parameter sets \((J_I,p)/\kappa_0\) of order \(10\), labeled by blue squares, lead to an exponential growth of \(\delta N_i\), as shown in Fig.~\ref{Fig4}(c). This growth indicates a breakdown of the perturbative approach rather than unbounded particle numbers in experiments.  A conservative conclusion is that experimentally accessible parameter regimes exist where the quantum corrections remain periodic and small compared to the background.

\section{Conclusions and Outlook}
We propose an experimentally feasible setup for a time crystal in coupled exciton-polariton condensates without coherent or periodic driving, distinguishing it from the widely studied Floquet time crystals in photonic cavities \cite{Seibold,Bakker,Li2024,Lee1,Niels,Walter,Weiss,Navarrete,Sonar2018}. Due to the large occupation numbers of the condensates, the master equation \eqref{modeu1} for the configuration in Fig.~\ref{Fig00} is numerically intractable using standard Fock-space methods \cite{qutip5}. Consequently, we analyze the system at the mean-field level and subsequently treat the quantum fluctuations perturbatively.


We have mapped the parameter space of the time-crystalline phase at the mean-field level, based on the conjecture that this regime corresponds to the intersection of the instability regions for all fixed points. This conjecture is supported by extensive numerical sampling, with no counterexamples identified. We analytically derived all fixed points of the SL equations \eqref{modeu1}, including the symmetric solutions \(N_{i\pm}^{\rm S}\) previously reported \cite{ARONSON1990403} and four previously unreported   asymmetric fixed points \(N_{i1,2,3,4}^{\rm AS}\). By evaluating the stability of these points via the Jacobian matrix, we identified the region in the three-dimensional \((\alpha,\beta,\gamma)\) parameter space that supports time crystals, yielding the necessary condition \(\alpha>\sqrt{5/4}\) for their emergence. Numerical bifurcation diagrams confirm continuous phase transitions between the symmetric, asymmetric, and time-crystalline states, in full agreement with our analytical findings.

We employ Bogoliubov perturbation theory to evaluate quantum fluctuations, where the resulting \(4\times4\) covariance matrix Eq.~\eqref{coveq} depends on eight parameters, while the mean-field dynamics involve only five. Due to this many-to-one mapping, distinct parameter sets corresponding to the same point in \((\alpha,\beta,\gamma)\) space can yield either divergent or stable oscillatory corrections (Fig.~\ref{Fig4}). Such divergences signal a breakdown of the perturbative expansion rather than physical singularities. Importantly, we identify regimes where quantum corrections remain stable and small relative to the mean-field background, ensuring the robustness of the time-crystalline phase.


Several directions merit further exploration. At the mean-field level, the analytical fixed-point solutions derived here offer a unified framework for characterizing the long-time dynamics of coupled SL equations, including amplitude death, oscillation death, and limit-cycle bifurcations phenomena that have largely been explored through numerical methods \cite{KOSESKA2013173,Xu2024,Koseska111}. Furthermore, it would be insightful to contrast our physical results with purely mathematical treatments of similar coupled systems \cite{Millan}. At the perturbative level, while we have derived the covariance matrix and demonstrated that both divergent and stable oscillatory corrections can emerge from the time-crystalline background, establishing a systematic criterion for their onset and completing a full stability analysis remain important open challenges.

\section*{Acknowledgments}
Hong-Jin Xiong and Xuan Ye contributed equally to this work and are considered co-first authors.
S.G. Acknowledges funding supports from the National Natural Science Foundation of China (Grant No. 12274034), the Guangdong Province Foreign Experts Project (Flexible Talent Introduction) Fund (No. 2025A1313010019), Shenzhen Specially Appointed Positions (No. 2025TC0140), and University Development Fund (No. UDF01003913). We thank Xing-Ran Xu and Zhen-Hua Guo for discussions related to this project.

\bibliographystyle{unsrt}  
\bibliography{Ref}         

\end{document}